\documentclass[amsmath,twocolumn,showpacs,amssymb,aps]{revtex4-1}
\usepackage{graphicx}
\usepackage{geometry}
\usepackage{subcaption}
\usepackage{pgfplots}
\usepackage{dcolumn}
\usepackage{bm}
\usepackage{slashed}
\usepackage{amsmath}
\usepackage{amssymb}
\usepackage{tikz}
\usepackage{bbold}
\usepackage{mathtools}
\usepackage{romannum}
\usepackage{natbib}
\usepackage{braket}
\newcommand{\cn}{{\mathcal C}_N}
\newcommand{\cnk}{{\mathcal C}_{Nk}}
\newcommand{\LE}{{\mathcal L}}
\newcommand{\LEk}{{\mathcal L}_k}
\begin{document}
\pagenumbering{arabic}
\title{Complexity, Information Geometry, and Loschmidt Echo near Quantum Criticality \\}
\author{Nitesh Jaiswal}
\email{nitesh@iitk.ac.in}	
\author{Mamta Gautam}
\email{mamtag@iitk.ac.in}
\author{Tapobrata Sarkar}
\email{tapo@iitk.ac.in}
\affiliation{
Department of Physics, Indian Institute of Technology Kanpur-208016, India}
\date{\today}
\begin{abstract}
We consider the Nielsen complexity ${\mathcal C}_N$, the Loschmidt echo ${\mathcal L}$, and the Fubini-Study complexity 
$\tau$ in the transverse XY model, following a sudden quantum quench, in the thermodynamic limit. 
At small times, the first two are related by ${\mathcal L} \sim e^{-{\mathcal C}_N}$. By computing a novel
time-dependent quantum information metric, we show that in this regime, ${\mathcal C}_N \sim d\tau^2$,
up to lowest order in perturbation. 
The former relation continues to hold in the same limit at large times, whereas the latter does not. 
Our results indicate that in the thermodynamic limit, the Nielsen complexity and the Loschmidt echo 
show enhanced temporal oscillations when one quenches from a close neighbourhood of the critical line, 
while such oscillations are notably absent when the quench is on such a line. We explain this 
behaviour by studying the nature of quasi-particle excitations in the vicinity of criticality. 
Finally, we argue that the triangle inequality for the Nielsen complexity might be 
violated in certain regions of the parameter space, and point out why one should be careful about the nature of the 
interaction Hamiltonian, while using this measure. 

\end{abstract}
\maketitle

\section{\label{Intro}Introduction}

Out of the many important aspects of quantum information theory, one which has been popular in the recent
past is that of complexity. The notion of complexity commonly appears in areas of computer science and 
computational complexity theories \cite{Arora, Moore}. There, complexity quantifies the difficulty to solve a particular task. 
In an informal description, this difficulty is the amount of computational resources required in the method 
of computation, such as the number of steps, or volume of memory, used to perform a sample task.
Of late, complexity has received attention in the context of widely different branches of physics \cite{Lloyd}. 
In fact, the recent flurry of activities in this field largely arose in the context of string theory and the
related gauge-gravity duality \cite{Myers1, Myers2}. The latter allows for computation of complexity in quantum field theories, which
may otherwise be intractable. Indeed, the current literature suggests that the notion of complexity is 
also related to deep questions in black hole physics, by the same notion of duality \cite{Susskind, Brown, Brown1}. 

What we are interested in here is a simpler situation, namely a many body quantum system 
with nearest neighbour interactions, in the thermodynamic limit, 
and we will mostly focus on the transverse field XY model. The notion of complexity 
is easier to understand here, and provides important physical insights. 
In quantum information theory, one quantification of complexity refers 
to a minimal number of universal or elementary gates required to 
construct a unitary transformation $U$, which acts on a given reference state  $\ket{\Psi_{R}}$ 
to produce a given target state $\ket{\Psi_{T}}$. Thus, with a few arbitrary choices (such as simple reference states), 
the NC is defined as the number of elementary gates in the optimal circuit. A natural difficulty with this
quantification arises, as the choices of good reference states and elementary gates have a theoretically 
infinite number of possibilities. Nielsen and collaborators \cite{Nielsen1, Nielsen2, Nielsen3} therefore
adapted a geometric method for the evaluation of complexity, dubbed the Nielsen complexity (NC) or the
circuit complexity. 
This geometric approach (explored in detail recently in \cite{Liu, xiong, Khan, tapo1}) 
involves a continuum of the unitary transformations $U(\tau)$ generated by any 
(possibly) time-dependent Hamiltonian $H(t)$, with $\tau$ parametrising a path in the Hilbert space. 

In this approach, one considers the trajectories in the space of unitaries such that the 
interesting trajectory satisfies $U(\tau=0)=\mathbb{1}$ and $U(\tau=1)=U$. Further, one identifies the optimal path 
or circuit, by minimising a cost functional defined for various possible paths. 
These cost functionals satisfy certain physicality conditions, and then these can be shown to define 
length functionals on a Riemannian manifold that arises in the space of unitary transformations, 
for a specific choice of the cost functional \cite{Myers1, Myers2}. The metric on the space of unitaries is written as 
\begin{eqnarray}
ds^{2} &=& 4\delta_{ab}\Big[\frac{1}{{\rm Tr} \big[J^{a}.J_{a}^{T}\big]}\text{Tr}
\Big(\partial_{\tau}{U(\tau)}\cdot U^{-1}(\tau)\cdot (J^{a})^{T}\Big)
\Big]\times\nonumber\\ &~&\Big[\frac{1}{{\rm Tr} \big[J^{b}.J_{b}^{T}\big]}\text{Tr}
\Big(\partial_{\tau}{U(\tau)}\cdot U^{-1}(\tau)\cdot (J^{b})^{T}\Big)\Big]^{*}~,
\label{met}
\end{eqnarray}
with $J^a$s denoting the generators of the underlying symmetry group with a superscript $T$ denoting 
the transpose, and $U$ denotes the unitary transformation
that takes a reference state to a target state. 
Then, the complexity is quantified by the minimal geodesic paths connecting $\ket{\Psi_{R}}$
and $\ket{\Psi_{T}}$. The complexity thus reduces to a variational computation to find geodesics on the 
space of unitary transformations. Namely, writing the metric of Eq. (\ref{met}) as $ds^{2}= g_{\alpha\beta}dy^{\alpha}dy^{\beta}$, where
$y^{\alpha}$ denotes a set of coordinates on the space of unitaries, the minimal (i.e., geodesic) distance on the space
of unitaries is the NC. 

A related quantity that is also of great interest is the Fubini-Study complexity (FSC) \cite{tapo1, FS}. 
Contrary to the NC, this quantification arises from the quantum information metric (QIM) 
\cite{Bengt, Provost, Polkov, Zanardi, tapo2, tapo3, Hamma, Venuti}, which is
the real part of a more generic structure called the quantum geometric tensor, whose imaginary part
gives the Berry phase. The QIM being the Riemannian metric induced on the parameter space of the Hamiltonian (which
will always be two dimensional in our case), measures the distance between two neighbouring quantum states. The FSC 
is then the geodesic distance on the parameter manifold. Specifically, given a wave function $\ket{\Psi}$, the 
geometric tensor reads
\begin{equation}
\chi_{ij}=\bra{\partial_{i}\Psi}\partial_{j}\Psi\rangle-\bra{
\partial_{i}\Psi}\Psi\rangle\bra{\Psi}\partial_{j}\Psi\rangle ~,
\label{qgt}
\end{equation}
with \(\partial_{i}\equiv\frac{\partial}{\partial\lambda^{i}}\), \(i=1,2,\cdots,m\), 
where \(m\) is the dimension of the parameter space ${\vec \lambda}$. The metric on the parameter manifold
$g_{ij}={\rm Re}[\chi_{ij}]$. Given the QIM, geodesic equations on the parameter manifold are computed
in a standard fashion but these often need to be solved numerically. If we parametrise the
geodesic distance by an affine parameter, then inverting the geodesic equation (possibly
numerically) gives a measure of the FSC. In this work, we will be interested in a 
quench situation, where the FSC becomes time dependent. 

Another important quantity which has been studied in great details in the literature for 
over more than a decade now, is the Loschmidt echo (LE), denoted here by $\LE$, 
originally introduced in the context of quantum chaos \cite{Peres}, and defined as
\begin{equation}
\LE = \big|\bra{\Psi_0}e^{iHt}e^{-iH_Ft}\ket{\Psi_0}\big|^2=\big|\bra{\Psi_0}e^{-iH_Ft}\ket{\Psi_0}\big|^2~,
\label{LEexp}
\end{equation}
where $\ket{\Psi_0}$ is an initially prepared ground state of the transverse XY model.
Here, $H$ is a time-independent Hamiltonian and $H_F = H + H_I$, with $H_I$ being a
perturbative interaction. The second expression in Eq. (\ref{LEexp}) follows as $\ket{\Psi_0}$ is
an eigenstate of $H$. The LE is in a broad sense a dynamical version of the 
QIM, and its decay and revival structures have received a lot of attention in the 
past decade \cite{Hamma1}, and will be of interest to us in this paper.

In \cite{tapo1}, the NC and FSC of a transverse XY spin chain was studied in a static situation, 
following a related work of \cite{Liu} in the context of the Kitaev model. 
It was shown in \cite{tapo1} that the derivative of the complexity
is a clear indicator of zero temperature quantum phase transitions (QPTs) in the XY model, in the sense that
it becomes divergent at the location of such transitions. Here, we first show some 
analytic results for the NC of the model, when the reference and target states are separated
by a small change in the parameters. This way, we are able to explicitly show that the
infinitesimal NC is related to the FSC, a formal argument for which was put forward in \cite{tapo1}. 
In fact, this allows us to check an important issue. Since the NC measures a distance, it
should follow the triangle inequality. What we find here is that this inequality is not 
satisfied even in the infinitesimal version of the NC, an issue that deserves further study
and will be commented upon here. 

We thereafter analyse a time-dependent sudden quench situation, with the XY model coupled to a central spin-1/2
system \cite{Ali, Ali1, Diptarka}. This was studied first by Zanardi and collaborators \cite{ZanLE} 
(see also \cite{LE1, LE2, LE3, LE4, LE5, LE6, AmitShraddhaVictor}) in the context of the LE. Here, we study this model in
the context of the NC and the FSC and compare these with the LE. 
Interestingly, in the thermodynamic limit, we show that for all times, a very generic formula that
holds in most part of the parameter space of the XY model is $\LE = e^{-\cn}$. While for small
times we find that $\LE \equiv e^{-d\tau^2}$, this relation is nonetheless challenged at
large times. In this paper, we further obtain the following results. 
a) While $\LE$ and the derivative of $\cn$ are indicators 
of quantum phase transitions for all times, $\tau$ only indicates such
transitions at large times. b) In the thermodynamic limit, $\LE$ and
$\cn$ show enhanced temporal oscillations when one quenches from the critical line, 
while such oscillations are completely absent when the quench is on such a line, and this 
is explained by studying the nature of quasi-particle excitations in the vicinity of this line. 
Finally, we also make some observations about the triangle inequality for $\cn$. We show
that this might be violated in certain cases and this indicates that the current study 
must be further pursued.

\section{Time-independent Nielsen complexity}
\label{static}

To set up the notations and conventions used in the rest of this paper, we will first
revisit some issues regarding the transverse XY model. 
We consider the one-dimensional spin-$1/2$ XY model in a transverse magnetic field, one
which exhibits quantum phase transitions. 
The Hamiltonian of the model reads
\begin{equation}
H=-\sum\limits_{l=-M}^{M}\!\!\left(\frac{1+\gamma}{4}\sigma^{x}_{l}\sigma^{x}_{l+1}
+\frac{1-\gamma}{4}\sigma^{y}_{l}\sigma^{y}_{l+1}-\frac{h}{2}\sigma^{z}_{l}\right)~,
\label{OriginalHamil}
\end{equation}
where $M=(N-1)/2$, for odd $N$, $\gamma$ is the anisotropy parameter, and $h$ the applied magnetic field,
and $\sigma$ denotes the Pauli matrices. Quantum phase transitions in this model occurs on the
lines $|h|=1$ and on $\gamma=0$.
To see this, the model is diagonalised using 
the Jordan-Wigner, Fourier and Bogoliubov transformations, and the energy eigenvalues are
\begin{equation}
\Lambda_{k\pm}=\pm\sqrt{(\cos k+h)^{2}+(\gamma\sin{k})^{2}},
\end{equation}
with $k=\frac{2\pi\lambda}{N}~,~~\lambda=-\frac{N-1}{2},....,-1,0,1,....,\frac{N-1}{2}~$.
The energy gap reads
\begin{eqnarray}
\Delta(k)=\Lambda_{k+}-\Lambda_{k-}
=2\sqrt{(\cos k+h)^{2}+(\gamma\sin{k})^{2}}~.\nonumber\\
\label{energygap}
\end{eqnarray}
The spectrum is gapless on the line $\gamma=0,~|h| \leq 1$, which signals an anisotropic 
transition line between two ferromagnetically ordered phases, and at $|h| = 1$ (for $k=0, \pi$), 
which are the Ising transition lines between a ferromagnetic and 
a paramagnetic phase. The ground state of the model is given by
\begin{eqnarray}
\ket{\Psi_0}_{h,\gamma}=\prod_{k>0}\bigg[\cos\left(\frac{\theta_{k}}{2}\right)\ket{0}_{k}\ket{0}_{-k}
&-&i\sin\left(\frac{\theta_{k}}{2}\right)\nonumber\\
&~&\times\ket{1}_{k}\ket{1}_{-k}\bigg]~,
\label{gsXY}
\end{eqnarray}
where $\ket{0}_k$ and $\ket{1}_k$ denote the vacuum and the single excitation states of Jordan-Wigner
fermions with momentum $k$. The Bogoliubov angle $\theta_k$ is given from
\begin{equation}
\cos\theta_{k}=\frac{\cos k+h}{\sqrt{(\cos k+h)^{2}+(\gamma\sin{k})^{2}}}~.
\label{Bogoliubov}
\end{equation}

As was derived in \cite{Liu} and extensively discussed in \cite{Liu}, \cite{tapo1}, the 
NC (denoted by ${\mathcal C}_N$) of a quadratic Hamiltonian is expressed in terms of the 
Bogoliubov angle $\theta_k$ that characterises its ground state, i.e.,
${\mathcal C}_N =\sum\limits_{k}^{}|\Delta\theta_k|^2~$, where 
$\Delta\theta_k = (\theta_k^T - \theta_k^R)/2$. Here, $\theta_k^R$ and $\theta_k^T$
are the Bogoliubov angles corresponding to reference and target states 
$\ket{\Psi_{R}}$ and $\ket{\Psi_{T}}$ that we have discussed above. 

We now write for the reference and the target states, $h^R=h$ and $h^T=h+\delta$, for 
a fixed non-zero value of the anisotropy parameter $\gamma$. Assuming
small $\delta$, we expand the NC up to a desired order in $\delta$, remembering that this 
series expansion is rendered invalid for regions of $h$ that are infinitesimally close to the
Ising phase transition. In the thermodynamic 
limit, it is then a fairly standard exercise to calculate the terms order by order in 
$\delta$ by going to the complex plane, and computing residues, where we replace
$\sum_k = N/(2\pi)\int_0^{\pi}dk$. Here and everywhere else, in the thermodynamic limit,
we will work with quantities per system size, and suppress the factor of $N$. 
Doing this, we finally obtain, 
\begin{eqnarray}
\cn\big|_{|h|<1} &=&\frac{\delta^2}{16|\gamma|\left(1-h^2\right)} 
+ \frac{h\delta^3}{16|\gamma|\left(1-h^2\right)^2} \nonumber\\
&+&\frac{\delta ^4 \left(7 \gamma ^2 \left(3 h^2+1\right)+h^2-1\right)}{384 |\gamma| ^3
\left(1-h^2\right)^3}+{\mathcal O}(\delta^5)\nonumber\\
\cn\big|_{|h|>1}&=& \frac{\gamma ^2 \delta ^2 |h|}{16 \left(h^2-1\right) {\mathcal A}^{3}}\nonumber\\
&\pm&\frac{\gamma ^2 \delta ^3 \left(\gamma ^2+4 h^4+\left(\gamma ^2-3\right) h^2-1\right)}{32
\left(h^2-1\right)^2 {\mathcal A}^{5}}+{\mathcal O}(\delta^4),~~~~~
\label{stat1}
\end{eqnarray}
with ${\mathcal A} = \sqrt{h^2+\gamma^2-1}$, and the $-(+)$ signs in the second term of 
the second equation above refers to the region $h>1 (<-1)$, respectively.  
The coefficients of $\delta^2$ in the first terms of the right hand side of the two equations in Eq. (\ref{stat1}) 
are the components of the information metric in the $h-\gamma$ plane \cite{Zanardi}, \cite{Polkov}. This is in lines with 
the argument given in \cite{tapo1}, i.e., the infinitesimal NC (with the differential element $dh$ identified with $\delta$) 
gives the line element on the parameter space. The other terms in this equation represent higher order corrections. 
Similarly, for a fixed value of the magnetic field $h$, writing $\gamma^R=\gamma$ and 
$\gamma^T=\gamma+\delta$, we obtain
\begin{eqnarray}
\cn\big|_{|h|<1} &=& \frac{\delta ^2}{16 |\gamma|  (|\gamma| +1)^2}
\mp\frac{(3 |\gamma| +1) \delta ^3}{32 \gamma ^2 (|\gamma| +1)^3}\nonumber\\
&+&\frac{(|\gamma|  (43 |\gamma| +28)+7) \delta ^4}{384 |\gamma| ^3 (|\gamma| +1)^4} + {\mathcal O}(\delta^5)~,
\label{stat2}
\end{eqnarray}
where the $-(+)$ sign in the second term has to be used for $\gamma>0(<0)$, respectively. 
The expression for $\cn\big|_{|h|>1}$ being too cumbersome to reproduce here, although we note 
that the ${\mathcal O}(\delta^2)$ term is indeed the $\gamma-\gamma$ component of the 
information metric listed in \cite{Zanardi}, \cite{Polkov}. 

A similar exercise can be done for the reference state being at $(h,\gamma)$ and the target state
at $(h+\delta, \gamma + \delta)$ where for the sake of simplicity, we choose the value of the
target parameters to be shifted by the same amount $\delta$. In this case also, the expression
for the complexity can be evaluated perturbatively, but are 
lengthy beyond the second order. We find that 
\begin{eqnarray}
\cn\big|_{|h|<1}= \frac{\delta ^2 \left(2-h^2+|\gamma|(|\gamma| +2)\right)}{16 |\gamma|  (|\gamma| +1)^2
\left(1-h^2\right)} + {\mathcal O}(\delta^3)~.
\label{stat3}
\end{eqnarray}
That Eqs. (\ref{stat1}) - (\ref{stat3}) give excellent approximations to the NC on the $h-\gamma$ 
parameter space except for points arbitrarily close to the critical lines, is readily checked,
by comparing with the numerically obtained values. 
Close to these lines however, the perturbative expansion in $\delta$ up to the first few orders that we
have done here is inadequate, and higher order terms start contributing significantly. One therefore
has to be careful in applications of the equations derived here, namely that the $\delta$-expansion 
should not be used to describe the physics of the cross-over regions. 

Now from the above discussion, we can examine the triangle inequality for the NC. Broadly speaking, since
the NC relates to a distance on a Riemannian manifold, the sum of the complexities resulting out of reaching
a target state from a given reference by two distinct operations should be greater than the complexity
of reaching the target by a single operation from the reference. We start with a given reference
state $(h,\gamma)$ and reach a target state $(h+\delta,\gamma + \delta)$. 
We can reach the target state by a combination of two operations ${\mathcal O}_1 : (h,\gamma) \to (h+\delta,\gamma)$
and ${\mathcal O}_2 : (h+\delta,\gamma) \to (h+\delta,\gamma+\delta)$. Alternatively, we can reach it by
a single operation ${\mathcal O} : (h,\gamma) \to (h+\delta,\gamma+\delta)$. The triangle inequality 
then implies that $\Delta = (\cn^{{\mathcal O}_1} + \cn^{{\mathcal O}_2})- \cn^{{\mathcal O}}\geq 0$
(for further discussions, see the recent article \cite{Yangetal}). We find that
\begin{eqnarray}
\Delta|_{|h|<1} &=& \pm\frac{\delta ^3}{32 \gamma ^2 \left(1-h^2\right)} \pm
\frac{\delta ^4 \left(8 \gamma  h- 3(1-h^2)\right)}{192 \gamma ^3 \left(1-h^2\right)^2}+{\mathcal O}(\delta^5)\nonumber\\
\Delta|_{|h|>1} &=& \pm\frac{\gamma\delta^2}{8{\mathcal A}^3}+{\mathcal O}(\delta^3)~,
\label{stat4}
\end{eqnarray}
where the $+(-)$ sign in the first equation has to be used for $\gamma>0(<0)$, respectively, 
and the $+(-)$ sign in the second equation has to be used for $h>1 (<-1)$, respectively. 
Thus, for $|h|<1$, at ${\mathcal O}(\delta^2)$, there is no contribution to $\Delta$. 
\begin{figure}[h!]
\centering
\includegraphics[width=0.4\textwidth, height=0.3\textwidth]{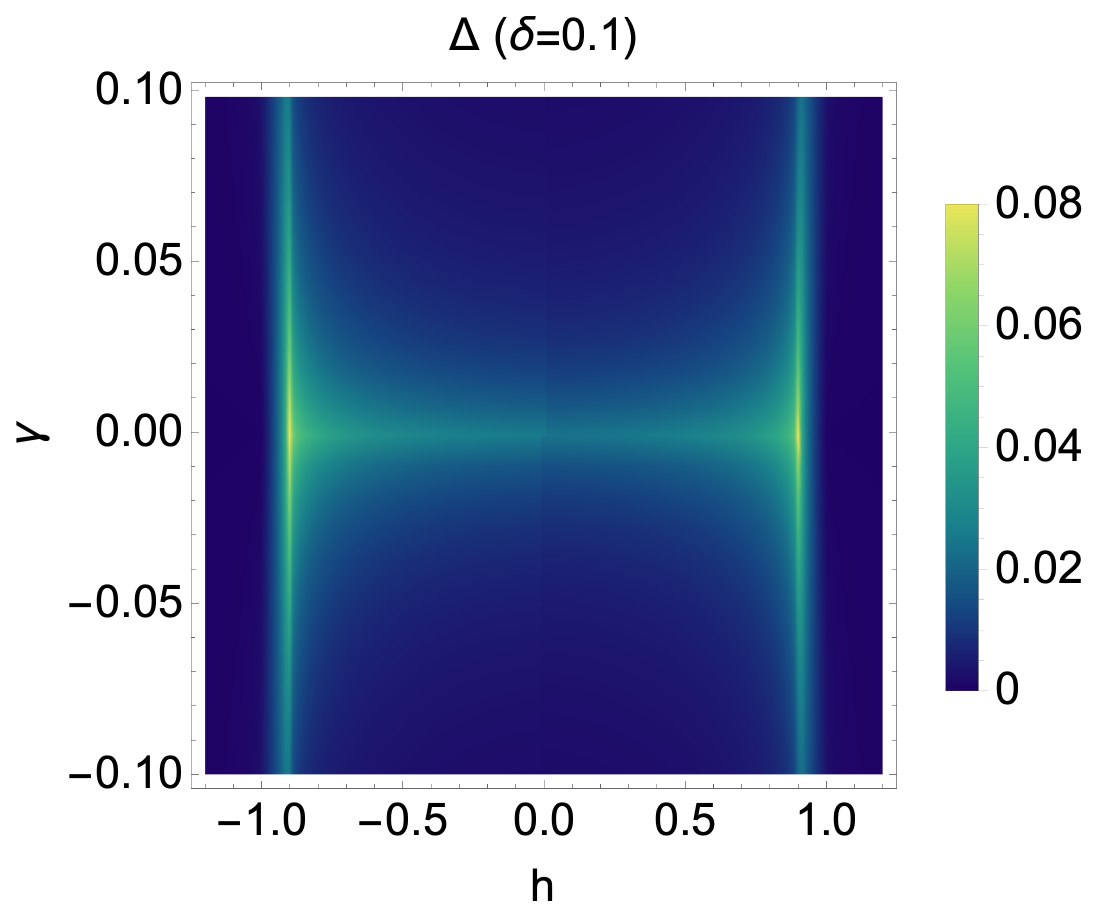}
\caption{$\Delta$ as a function of $h$ and $\gamma$ with $\delta = 0.1$}
\label{HM1}
\end{figure}
However, as one approaches the critical lines $|h|=1$, the situation changes and $\Delta$ starts getting
contributions from ${\mathcal O}(\delta^3)$ terms, enhanced by the $\sim (1-h)$ terms in the denominator 
of Eq. (\ref{stat4}). There is however no divergence here, as the power of the infinitesimals will always
be non-negative, as is apparent from this equation. Let us focus on the region $\gamma>0$, and take $\delta>0$.
With these assumptions, for $h-1\sim {\mathcal O}(\delta)$ upto
which our series expansion is maximally valid, $\Delta|_{|h|<1}$ is positive definite, as should be the
case. However, we find that $\Delta|_{h<-1}$ becomes negative. The triangle inequality
seems to be violated here. To avoid this, we can define the target state as 
$(h+{\rm sign}(h)\delta,\gamma + {\rm sign}(\gamma)\delta)$, in which case the violations
of the triangle inequality is avoided.

Doing this, the overall picture is that whereas deep in the 
ferromagnetically or paramagnetically ordered phases, infinitesimal $\Delta$ is always close to zero,
it picks up a finite value when the path is close to a critical point, the effect being enhanced when the
multi-critical point is approached. All these facts are illustrated in Figure (\ref{HM1}), where
we have defined the target states along with the signs of the parameters as discussed above, and by
numerical integration, we show $\Delta$ on the $h-\gamma$ plane in the thermodynamic limit, with $\delta = 0.1$. 

\section{Quench scenarios}
\label{quench}

We now consider the XY model with a quantum quench, assuming the sudden quench approximation. 
Here, following the original work of \cite{ZanLE}, we consider an XY model environment 
coupled to a two-level central spin-1/2 system. Due to this coupling, the wave function of the
XY model evolves in two distinct branches, as the central spin backreacts differently on the
environment depending on whether it is in the ground state $\ket{g}$ or the excited state $\ket{e}$. 
We will take the interaction Hamiltonian to be $H_I$, so that the total Hamiltonian is $H_F=H+H_I$,
with $H$ given in Eq. (\ref{OriginalHamil}), and 
\begin{eqnarray}
H_I &=& c_1\frac{\delta}{2}\ket{e}\bra{e}\sum\limits_{l=1}^{N}\sigma^{z}_{l} \nonumber\\
&-& c_2\frac{\delta}{4}\ket{e}\bra{e}\sum\limits_{l=1}^{N}
\left(\sigma^{x}_{l}\sigma^{x}_{l+1}-\sigma^{y}_{l}\sigma^{y}_{l+1}\right)~.
\label{GenHamil}
\end{eqnarray}
The first coupling term which corresponds to a transverse quench was studied in \cite{ZanLE} and the 
second one, corresponding to an anisotropic quench was elaborated upon in \cite{AmitShraddhaVictor}.
Here, $c_1, c_2=0,1$ correspond to turning the coupling(s) off or on respectively, and we have taken the same interaction
strength $\delta$, which simplifies the computations considerably while bringing out the essential physics. 

By a fairly standard approach, one first writes the ground state $\ket{\Psi_0}_{h,\gamma}$ 
of $H$ in terms of that of $H_F$, labeled $\ket{\Psi_{0,k}}_{h+c_1\delta,\gamma+c_2\delta}$ for the $k$th
Fourier mode. This gives
\begin{equation}
\ket{\Psi_0}_{h,\gamma}=\prod_{k>0}\left[\cos\Omega_k - i\sin\Omega_k\chi_{k}^{\dagger}
\chi_{-k}^{\dagger}\right]\ket{\Psi_{0,k}}_{h+c_1\delta,\gamma+c_2\delta}~,
\end{equation}
where we have defined 
\begin{eqnarray}
\Omega_k &=& \frac{1}{2}\left[\theta_k(h,\gamma) - \theta_k(h+c_1\delta,\gamma + c_2\delta)\right]\nonumber\\
\chi_k &=& \cos\left(\frac{\theta_k(h+c_1\delta,\gamma+c_2\delta)}{2}\right)c_k \nonumber\\
&+&i\sin\left(\frac{\theta_k(h+c_1\delta,\gamma+c_2\delta)}{2}\right)c_{-k}^{\dagger}~,
\end{eqnarray}
with the operators $c_k$ and $c_{-k}^{\dagger}$ being the Fourier operators, and 
$\theta_{k}$ is the Bogoliubov angle defined in Eq. (\ref{Bogoliubov}) with the arguments
appropriately defined. 

To compute the complexity, the reference and the target states are chosen to be $\ket{\Psi_0}_{h,\gamma}$
and $\ket{\Psi_{e}(t)}$ respectively, where, for a transverse quench with $c_{1}=1$ and $c_2=0$ for example,
$\ket{\Psi_{e}(t)}=e^{-iH_{F}(h+\delta,\gamma)t}\ket{\Psi_0}_{h,\gamma}$, i.e.,
\begin{eqnarray}
\ket{\Psi_{e}(t)}&=&\prod\limits_{k>0}\!e^{-i\epsilon_{k}(h+\delta,\gamma)t}\bigg[\cos(\Omega_{k})
-i\sin(\Omega_{k})\times\notag\\
& &\chi_{k}^{\dagger}\chi_{-k}^{\dagger}e^{-2i\epsilon_{k}(h+\delta,\gamma)t}\bigg]\ket{\Psi_{0,k}}_{h+\delta,\gamma}.
\label{wavefn}
\end{eqnarray}
The computation of the NC is now standard, and was outlined in \cite{Liu}. The final result for
a general quench is  
\begin{equation}
\cn(t)\equiv \sum\limits_{k}{\mathcal C}_{Nk}=\sum\limits_{k}\Phi_{k}^{2}(h+c_1\delta,\gamma+c_2\delta,t)~,
\label{compt}
\end{equation}
where \begin{equation}
\Phi_{k}=\arccos\left(\sqrt{1-\sin^{2}(2\Omega_{k})\sin^{2}(\epsilon_{k}(h+c_1\delta,\gamma+c_2\delta)t)}\right)~,
\label{Phik}
\end{equation}
with the single particle excitations 
$\epsilon_k(h+c_1\delta,\gamma+c_2\delta) = \sqrt{(h+c_1\delta+\cos k)^2 + (\gamma+c_2\delta)^2\sin^2k}$. 
We also record the expression for the Loschmidt echo $\LE$ given from Eq. (\ref{LEexp}).
Noting that $\ket{\Psi_0}_{h,\gamma}$ is an eigenket of $H$, we get $\log\LE=\sum_k {\mathcal L}_k$, with
\begin{equation}
{\mathcal L}_k = \log\left(1-\sin^{2}(2\Omega_{k})\sin^{2}(\epsilon_{k}(h+c_1\delta,\gamma+c_2\delta)t)\right)~.
\label{ellk}
\end{equation}
Evidently then, at small times, by expanding the arccosine and the logarithm, we have the relation
\begin{equation}
\LE \simeq e^{-\cn}~,~~t\to 0~.
\label{LENC}
\end{equation}
We now illustrate this with the transverse quench. 

\subsection{Transverse Quench : Complexity at small times}
\label{compsmallt}

First, we consider a transverse quench, obtained by setting $c_1=1$ and $c_2=0$ in Eq. (\ref{GenHamil}). 
Here, for small $t$ and small $\delta$, we proceed by expanding the complexity of Eq. (\ref{compt}) in an
appropriate power series. Due to the nature of the terms involved, a controlled expansion (in terms of 
a single small parameter) is not possible here, and the two small parameters $t$ and $\delta$ come with unequal
powers. Hence we retain smallest powers in both. Then, by  following the same procedure as outlined in section (\ref{static}),
we get up to low orders in $\delta$ and $t$,
\begin{eqnarray}
\cn\big|_{|h|<1} &=& \cn^{reg}+ \frac{|\gamma|\delta ^2 t^2}{2\left(|\gamma| +1\right)}-
\frac{h \gamma ^2 \delta ^3  t^4}{3 (1+|\gamma|)^2}
~,\nonumber\\
\cn\big|_{|h|>1} &=& \cn^{reg}+\frac{\gamma ^2 \delta ^2 t^2 
\left(|h|-{\mathcal A}\right)}{2{\mathcal A}\left(1-\gamma ^2\right)}\nonumber\\
&\pm& \frac{\gamma ^2 \delta ^3 t^4 \left(\gamma^2-|h|\left(\gamma ^2+1\right) 
\left({\mathcal A}-|h|\right)-1\right)}
{3{\mathcal A}\left(1-\gamma^2\right)^2}~.\nonumber\\
\label{cnapprox}
\end{eqnarray} 
These are also the analytic expressions for $(-\log\LE)$ in the small $t$ limit, in
the mentioned regions of $h$. 
Here, the $+$ sign in the second line of the 
second equation above is appropriate for $h>1$ while in the region $h<-1$, one requires
to use the $-$ sign. The higher orders terms in these equations are lengthy, and we omit them for brevity. 
Also, the ``regular'' term in the expressions (i.e., ones
that do not have any pole in the complex plane and are evaluated directly) is given by
$\cn^{reg} = -\frac{1}{12}\gamma ^2 \delta ^2 t^4$
to the order that we are considering. 
\begin{figure}[h!]
\centering
\includegraphics[width=0.4\textwidth, height=0.3\textwidth]{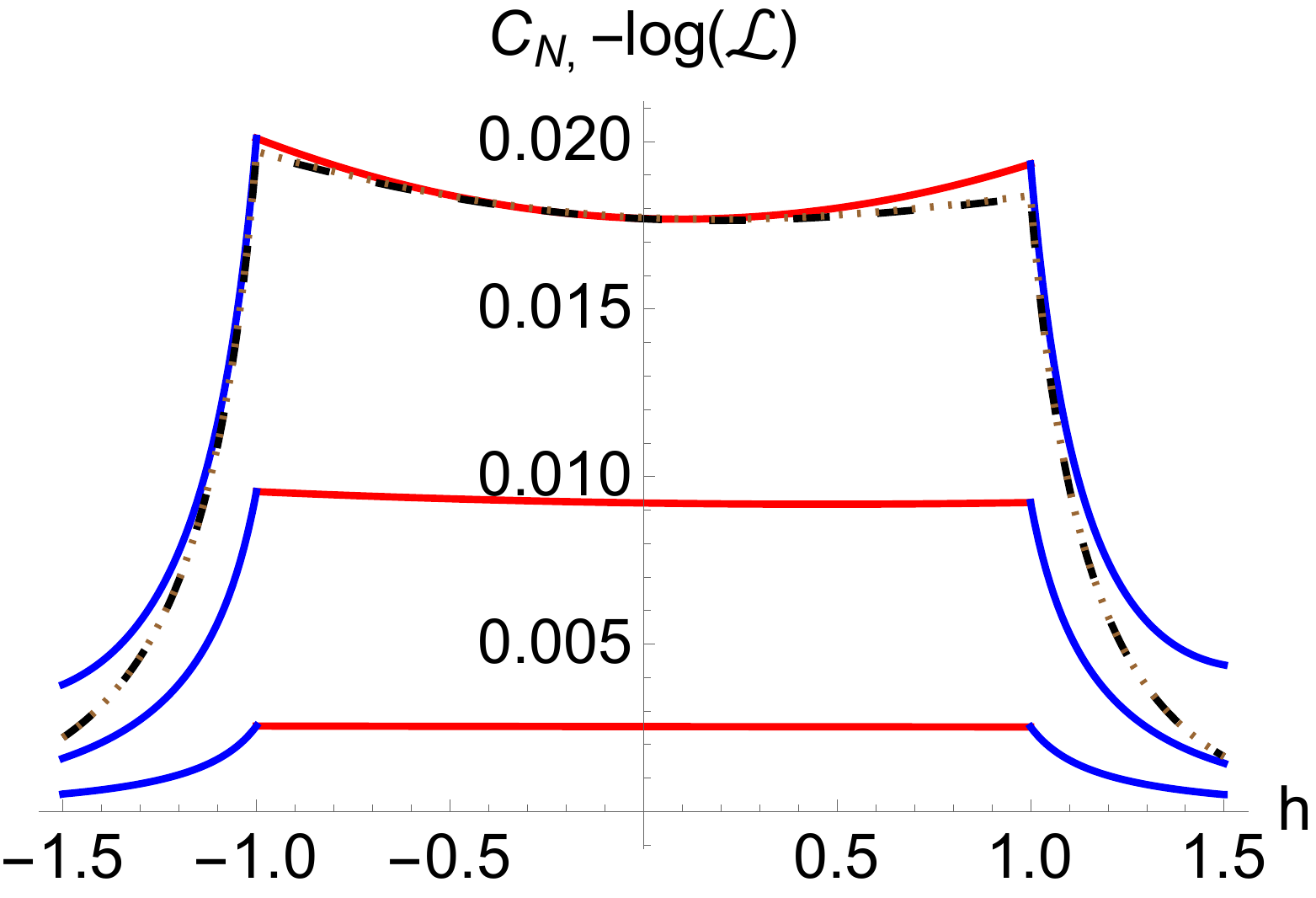}
\caption{$\cn$ and $(-\log\LE)$ as a function of $h$ for a transverse quench with $\gamma=0.5$ and $\delta = 0.1$ for 
$t=0.5$ (bottom), $1$ (middle) and $1.5$ (top). The numerically computed $\cn$
at $t=1.5$ is shown by the dashed black line, from which $(-\log\LE)$
shown by the brown dotted line (at $t=1.5$) is indistinguishable.}
\label{compsmallt1}
\end{figure}

In Fig. (\ref{compsmallt1}), we show $\cn$ computed from Eq. (\ref{cnapprox}) 
as a function of $h$ for $\gamma=0.5$ and $\delta = 0.1$,
for three different values $t=0.5$ (bottom), $t=1$ (middle) and $t=1.5$ (top). 
The numerically computed values are indistinguishable for the first two values of $t$, and
starts differing from the approximate value of Eq. (\ref{cnapprox}) only for larger
values of the time, where is shown by the dashed black curve, which depicts the numerical value
of $\cn$ computed from Eq. (\ref{compt}) at $t=1.5$ for $(\gamma,\delta)=(0.5,0.1)$, in
the thermodynamic limit. The dotted brown line in the figure corresponds to 
$(-\log\LE)$ computed numerically in the thermodynamic limit at $t=1.5$, 
which is indistinguishable from $\cn$, confirming Eq. (\ref{LENC}). For 
values of $t$ smaller than this, $(-\log\LE)$ is indistinguishable from 
$\cn$ computed via Eq. (\ref{cnapprox}). 

\begin{figure}[h!]
\centering
\includegraphics[width=0.4\textwidth, height=0.3\textwidth]{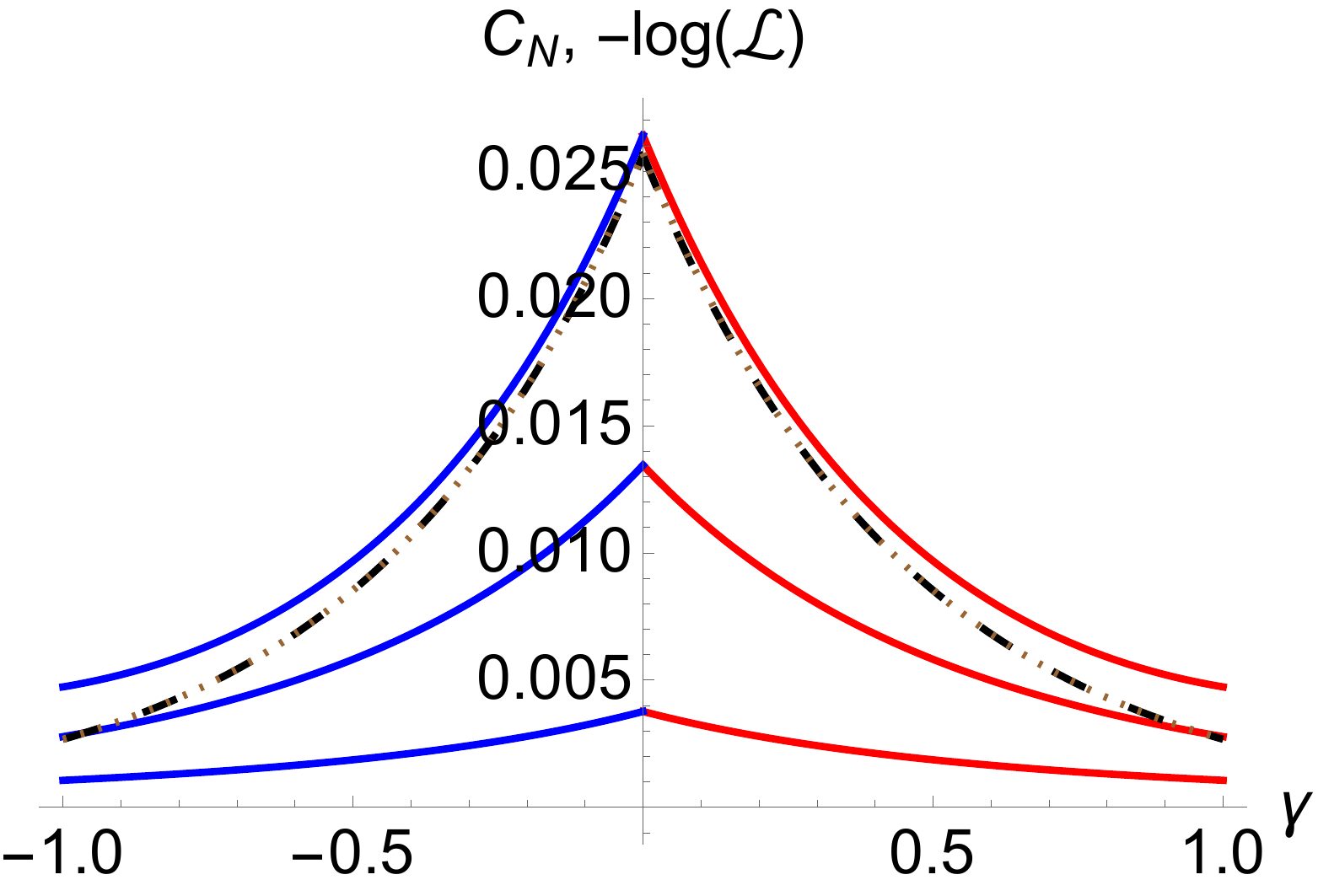}
\caption{$\cn$ and $(-\log\LE)$ as a function of $\gamma$ for an anisotropic quench with $h=0.5$ and $\delta = 0.1$ for 
$t=0.5$ (bottom), $1$ (middle) and $1.5$ (top). The numerically computed $\cn$
at $t=1.5$ is shown by the black dashed line, which is indistinguishable from 
$-(\log\LE)$ shown in dotted brown.}
\label{compsmallt2}
\end{figure}

A similar analysis holds for the anisotropic quench, and we have left the details 
to Appendix \ref{AppendixA}. To compare with the transverse quench, we present 
the behaviour of the $\cn$ and $(-\log\LE)$ with $\gamma$ in Fig.(\ref{compsmallt2}) for the anisotropic 
quench case, using Eq. (\ref{aniso1}). The solid lines in this figure correspond
to $t=0.5$ (bottom), $t=1$ (middle) and $t=1.5$ (top). As before, we find that
the numerically evaluated value of $\cn$ and $(-\log\LE)$ here are indistinguishable at small
values of $t$ with the ones computed via Eq. (\ref{aniso1}), and that 
the difference coming for larger values, depicted by the dashed 
black line corresponding to $t=1.5$. The dotted brown line in this figure indicates
$(-\log\LE)$ which is again indistinguishable from $\cn$, as was the case with the
transverse quench. 

For both the transverse and the anisotropic quench schemes, it is clear from Figures
(\ref{compsmallt1}) and (\ref{compsmallt2}) that the NC, and the LE
shows their nonanalytical nature at the critical points i.e., $|h|=1$ and $\gamma=0$. As in the
time-independent case, this is how $\cn$ is indicative of the zero temperature 
quantum phase transitions at the Ising and the critical anisotropy lines. We will now
tie up this result with the QIM. 

\subsection{Transverse Quench : QIM at small times}
\label{imsmallt}

As we have mentioned in the beginning of this section, the quench scenario that we consider here is 
obtained by coupling the transverse XY model to a central spin $1/2$ system. Once the
interaction Hamiltonian is turned on, an initially prepared ground state of the 
XY model $\ket{\Psi_0}_{h,\gamma}$ evolves in two branches, which we denote by $\ket{\Psi_g(t)}=
e^{-iHt}\ket{\Psi_0}_{h,\gamma}$ and $\ket{\Psi_e(t)}=e^{-iH_F t}\ket{\Psi_0}_{h,\gamma}$. The full wave function 
can then be written as 
\begin{equation}
\ket{\Psi(t)}= k_g{\ket g}\otimes\ket{\Psi_g(t)} + k_e{\ket e}\otimes\ket{\Psi_e(t)}~,
\label{genket}
\end{equation}
where the central two-level system is initially prepared in the normalised state $k_g\ket{g}
+k_e\ket{e}$. The precise values of the coefficients $k_g$ and $k_e$ (with $|k_g|^2+|k_e|^2=1$) are unimportant as far as
the NC and the LE is concerned. However, if we want to find the QIM corresponding to the
full wavefunction, we find using Eq. (\ref{qgt}) and
Eq. (\ref{genket}) that 
\begin{equation}
g_{ij} = |k_g|^2g^g_{ij} + |k_e|^2g^e_{ij} + |k_gk_e|^2{\mathcal A}_i{\mathcal A}^*_j~,
\label{branches}
\end{equation}
where ${\mathcal A}_i=\langle\partial_i\Psi_g(t)|\Psi_g(t)\rangle - \langle\partial_i\Psi_e(t)|\Psi_e(t)\rangle$.
Here, $g^g_{ij}$ and $g^e_{ij}$ are the metrics computed out of only the ground or only the excited state wavefunctions
in Eq. (\ref{genket}), respectively. For $\delta = 0$, ${\mathcal A}_i=0$ and we readily see that 
the QIM reduces to that of the ground state of the transverse XY model. However, for non-zero $\delta$, the
situation is more intricate, although even in that case, in a perturbative expansion in $\delta$, 
the ${\mathcal O}(\delta^0)$ term in $g^e_{ij}$ combines with $g^g_{ij}$ to produce the ground state QIM. 

Now in order to compare with the NC or the LE, the full wavefunction is less useful, and we will
need to focus on the QIM computed out of $\ket{\Psi_e(t)}$. This is because as we have already discussed,
as far as the NC is concerned, the reference and target states are taken to be $\ket{\Psi_0}_{h,\gamma}$ and $\ket{\Psi_e(t)}$. 
Even if we took the reference state to be $\ket{\Psi_g(t)}$, the NC (Eq. (\ref{compt})) remains unchanged, 
as $\ket{\Psi_0}_{h,\gamma}$ and $\ket{\Psi_g(t)}$ are related only by a phase. To contrast the FSC with the NC and LE, 
it is then more appropriate to compute the QIM corresponding to $\ket{\Psi_e(t)}$. Indeed, as we will see now,
this provides interesting insights. 

Of course, one might wonder that for the transverse quench, since $\gamma$ is held fixed, one would
not obtain a meaningful metric tensor, since $h$ is the only variable. Interestingly, in this case, we 
obtain a time component of the metric. For the QIM of the ground state of the transverse XY model 
or for the time evolved ground state, this is absent 
essentially due to the same argument as above : the time evolution of $\ket{\Psi_0}_{h,\gamma}$ with $H$ simply
adds a constant phase factor. For $\ket{\Psi_e(t)}$, the situation is more non-trivial, and gives
rise to a $t-t$ component of the metric. The metric is then meaningful in the $t-h$ plane. 

We compute the information metric close to $t=0$, and for small $\delta$. 
Using Eq. (\ref{wavefn}) in Eq. (\ref{qgt}), we find here 
after an elaborate but straightforward computation that there is a $\delta=0$ component of
the QIM which reduces to the QIM of the ground state of the transverse XY model reported
in \cite{Zanardi}, \cite{Polkov} (of which only the $g_{hh}$ component is relevant here, from the arguments
above). This is more appropriately related to $g^g_{ij}$ as discussed after Eq. (\ref{branches}). 
The components proportional to powers of $\delta$ are the quantities of interest here and we 
present them below (we drop the superscript $e$ in the expressions below to avoid cluttering of notation). 
We get, up to the lowest orders in $t$ and $\delta$,
\begin{eqnarray}
g_{tt}\big|_{|h|<1} &=& \frac{|\gamma|\delta^2}{2\left(1+|\gamma|\right)}~,\nonumber\\
g_{tt}\big|_{|h|>1} &=& \frac{\gamma ^2 \delta ^2 
\left(|h|-{\mathcal A}\right)}{2{\mathcal A}\left(1-\gamma ^2\right)}~,\nonumber\\
g_{hh}\big|_{|h|<1}&=& -
\frac{h \gamma ^2 \delta  t^4}{3 (1+|\gamma|)^2}~,\nonumber\\
g_{hh}\big|_{|h|>1}&=& \pm\frac{\gamma ^2 \delta t^4 \left(\gamma^2-|h|\left(\gamma ^2+1\right) 
\left({\mathcal A}-|h|\right)-1\right)}
{3{\mathcal A}\left(1-\gamma^2\right)^2}~,~~\nonumber\\
g_{th}&=&\frac{t}{\delta}g_{tt}+\frac{\delta}{t}g_{hh}-\frac{1}{6}\gamma ^2 \delta  t^3~,
\label{tqim}
\end{eqnarray}
where the $+(-)$ sign in the fourth equation refers to the region $h>1 (h<-1)$, respectively,
and the expression for $g_{th}$ is valid for both the $|h|<1$ and $|h|>1$ regions. 
A similar analysis holds for the anisotropic quench, and we have left the details to Appendix \ref{AppendixA}.
Note that the expression for $g_{tt}$ is exact. As can be checked from Eq. (\ref{qgt}),
there are no corrections to this expression beyond ${\mathcal O}(\delta^2)$. 

From Eq. (\ref{cnapprox}) and Eq. (\ref{tqim}), we readily 
see that the lowest order terms in $\cn$ and $d\tau^2$ is 
${\mathcal O}(\delta^2t^2)$, and at this order, we have $\cn \sim d\tau^2$, once we 
identify in the line element $dh\sim\delta$ and $dt\sim t$ (as appropriate for small times). 
Hence ${\mathcal L} = e^{-d\tau^2}$ for small times and to lowest order in the perturbing parameter $\delta$.
The more general relation valid up to the order that we consider in Eqs. (\ref{cnapprox}) and (\ref{tqim}) is
given as $3\cn =d\tau^2 - \cn^{reg}$ where the last term was defined after Eq. (\ref{cnapprox}).
That this last relation holds in the case of the anisotropic quench as well can be checked from the
formulae presented in Appendix \ref{AppendixA}.

Whereas the relation between $\cn$ and $\LE$ essentially followed from their form at small times, the 
outlined relation with the line element of the QIM does not. Indeed, for the time-independent case, the
relation $\cn \sim d\tau^2$
between the NC and the QIM follows by definition of $\cn$ \cite{tapo1}. However, in the present situation, one 
cannot possibly guess this relation from the definition of $\cn$ in Eq. (\ref{compt}). Quite
surprisingly, the $t-t$ component of the QIM plays a crucial role here. 
The fact that the three
fundamental quantities used in the study of quantum information theory are related by a simple 
formula is indeed quite striking. We have established this here for a specific case of the transverse
XY model.  

\subsection{Transverse Quench : FSC at small times}
\label{fscsmallt}

Having obtained the QIM, we now investigate the FSC for the transverse quenched model. 
We need to carefully clarify the meaning of the FSC here. As we have already mentioned,
the QIM computed from $\ket{\Psi_e(t)}$ also ``contains'' the static case, namely the ground state 
QIM appears at $\mathcal{O}(\delta^0)$. However, this is more appropriately 
associated with the branch of the XY model that couples with the ground state of the 
central spin half system. Hence, it makes sense to study that part of metric which is
dependent on $\delta$ and reflects the true effects of the coupling of the XY model to the 
central spin. What we need to do here is to compute $\tau$ as a function of the 
model parameters, and it will be enough for us to focus on the region $|h|<1$ to 
illustrate our point. 
We consider three cases, with $\gamma = 0.5$, and take the initial value of the time
to be $0.01$. In the first case, we consider $h=0.88, \delta=0.1$, the second case
being $h=0.90, \delta=0.05$, and the third is $h=0.92, \delta=0.01$. In all cases, we
numerically solve the two geodesic equations arising from Eq. (\ref{tqim}) with appropriate
boundary conditions, which involve the initial values of $t$ and $h$, and their derivatives
with respect to the affine parameter. While in all cases we fix the initial value
$dh/d\tau = -0.1$, the initial value of $dt/d\tau$ is fixed from the normalisation condition
$g_{ij}(dx^i/d\tau)(dx^j/d\tau)=1$. This then determines the geodesic evolution of $h$ with time.
Now having obtained the numerical solution of the geodesic
equations, we invert them using a standard root finding procedure in Mathematica. This will
then give us the solution of the FSC as a function of $h$, up to the phase boundary. 
\begin{figure}[h!]
\centering
\includegraphics[width=0.4\textwidth, height=0.3\textwidth]{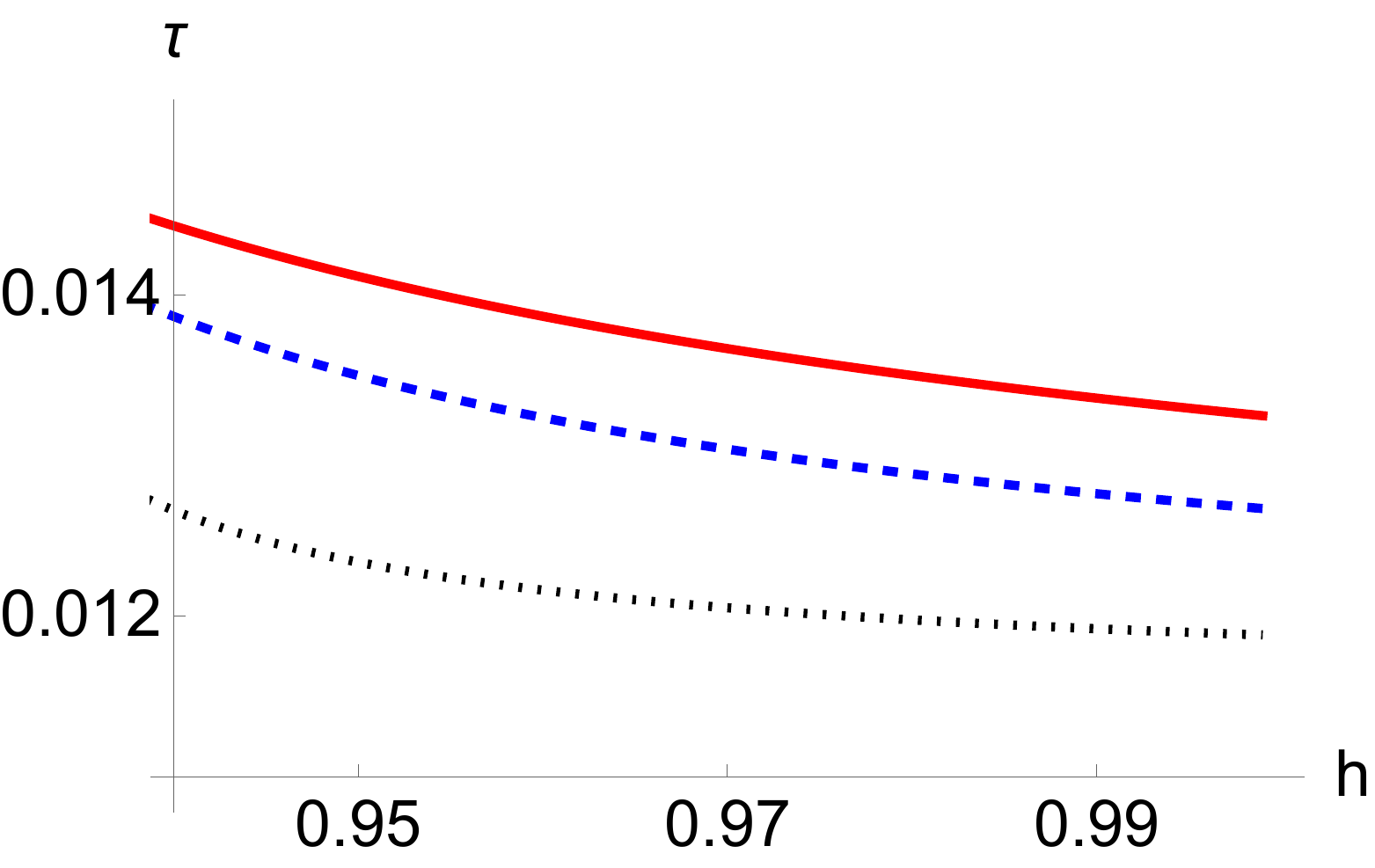}
\caption{The affine parameter $\tau$ as a function of $h$ for initial values $h=0.88$ (solid red), 
$h=0.90$ (dashed blue) and $h=0.92$ (dotted black).}
\label{geo2}
\end{figure}
Fig. (\ref{geo2}) shows the affine parameter $\tau$ as a function of $h$, with 
the initial values chosen as $h=0.88$ (solid red), $h=0.90$ (dashed blue) and $h=0.92$ (dotted black). Clearly, there is
no special behaviour of $\tau$ as we reach the phase boundary $h=1$. 
The reason for this is clear. The metric for the excited state wavefunction at small times is regular 
throughout the region $|h|<1$ and so is the Ricci scalar computed out
of this metric. The $(t,h)$ parameter manifold is thus divergence-free and
this information is reflected in the behaviour of the geodesics. 

\section{Transverse Quench : Complexity at finite times}
\label{transversefinite}

We will now consider the NC and the LE for a transverse quench at finite times, given from
Eqs. (\ref{compt}) and (\ref{ellk}), where we turn off the perturbation on $\gamma$. The analysis 
of $\cn$ and $\LE$ become complicated here, due to the nature of the expressions involved. 
However, we can make the following statements in momentum space. 
\begin{figure}[h!]
\centering
\includegraphics[width=0.4\textwidth, height=0.3\textwidth]{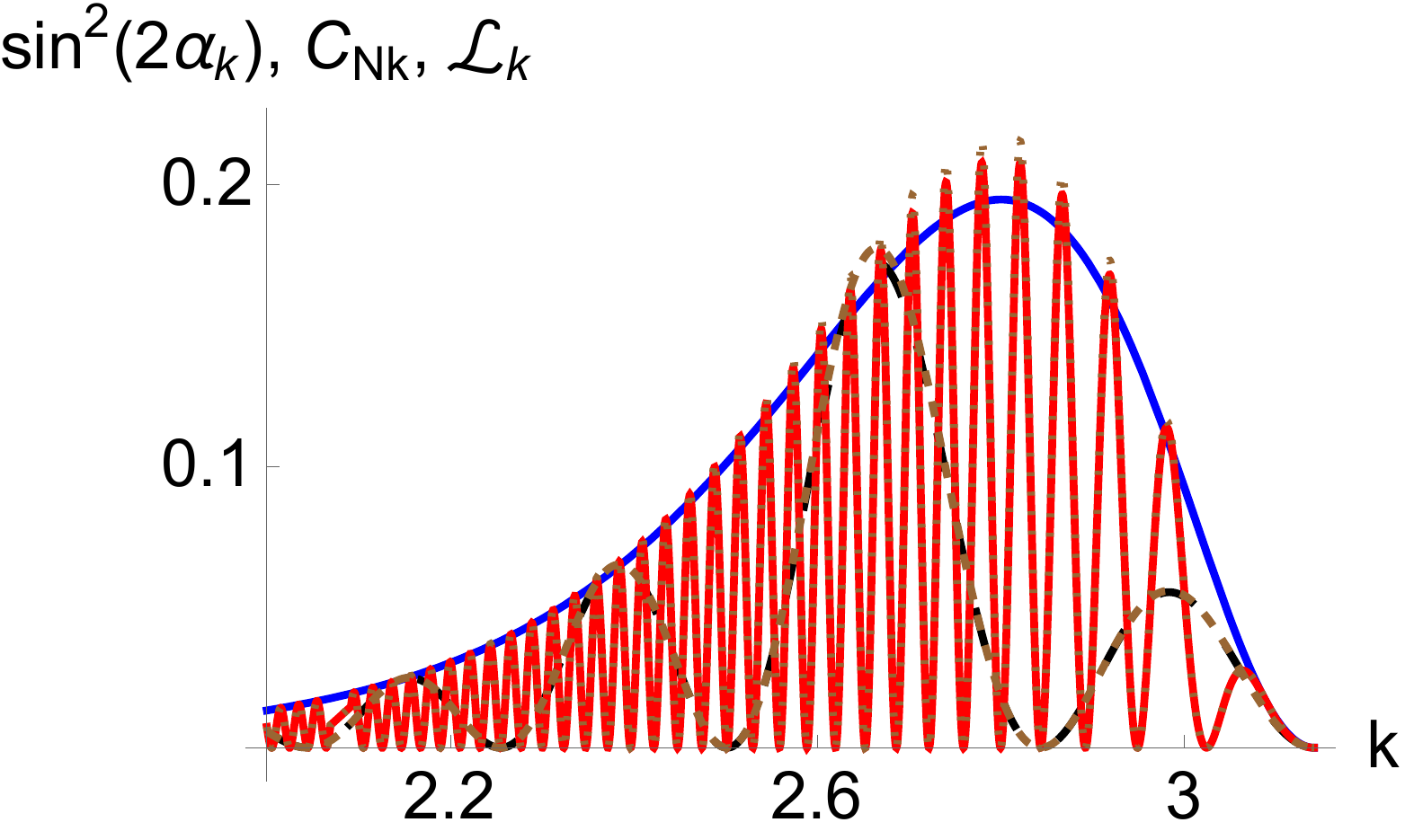}
\caption{$\sin^2(2\alpha_k)$ (solid blue), $\cnk$ at $t=20$ (dashed black) and 
$\cnk$ at $t=200$ (solid red) as a function of $k$ for $h=0.8, \gamma=0.5, 
\delta = 0.1$. $\LEk$ is plotted in dashed brown and is indistinguishable 
from the $\cnk$ graphs.}
\label{compL1}
\end{figure}
First, we note that due to the nature of $\Omega_k$ and $\epsilon_k$ given in 
Eq. (\ref{compt}), ${\mathcal C}_{Nk}$ and ${\mathcal L}_k$ are oscillatory functions of $k$, 
with the oscillation amplitude controlled by the $\sin^2(2\alpha_k)$, which acts as a modulation
function, away from the Ising transition lines $h=\pm 1$. We have depicted this
in Fig. (\ref{compL1}), where we have taken $h=0.8,\gamma=0.5$ and $\delta = 0.1$.
Here, the solid blue line represents $\sin^2(2\alpha_k)$. The black dashed
line gives $\cnk$ at $t=20$ while the solid red line represents $\cnk$ at $t=200$.
In this figure, we have also plotted $\LEk$, which are indistinguishable from 
the $\cnk$ lines.  
From the figure, we see that while a single Fourier mode contributes maximally
to $\cnk$ and $\LEk$ for small but finite times, several modes start contributing as the 
time increases. 
\begin{figure}[h!]
\centering
\includegraphics[width=0.4\textwidth, height=0.3\textwidth]{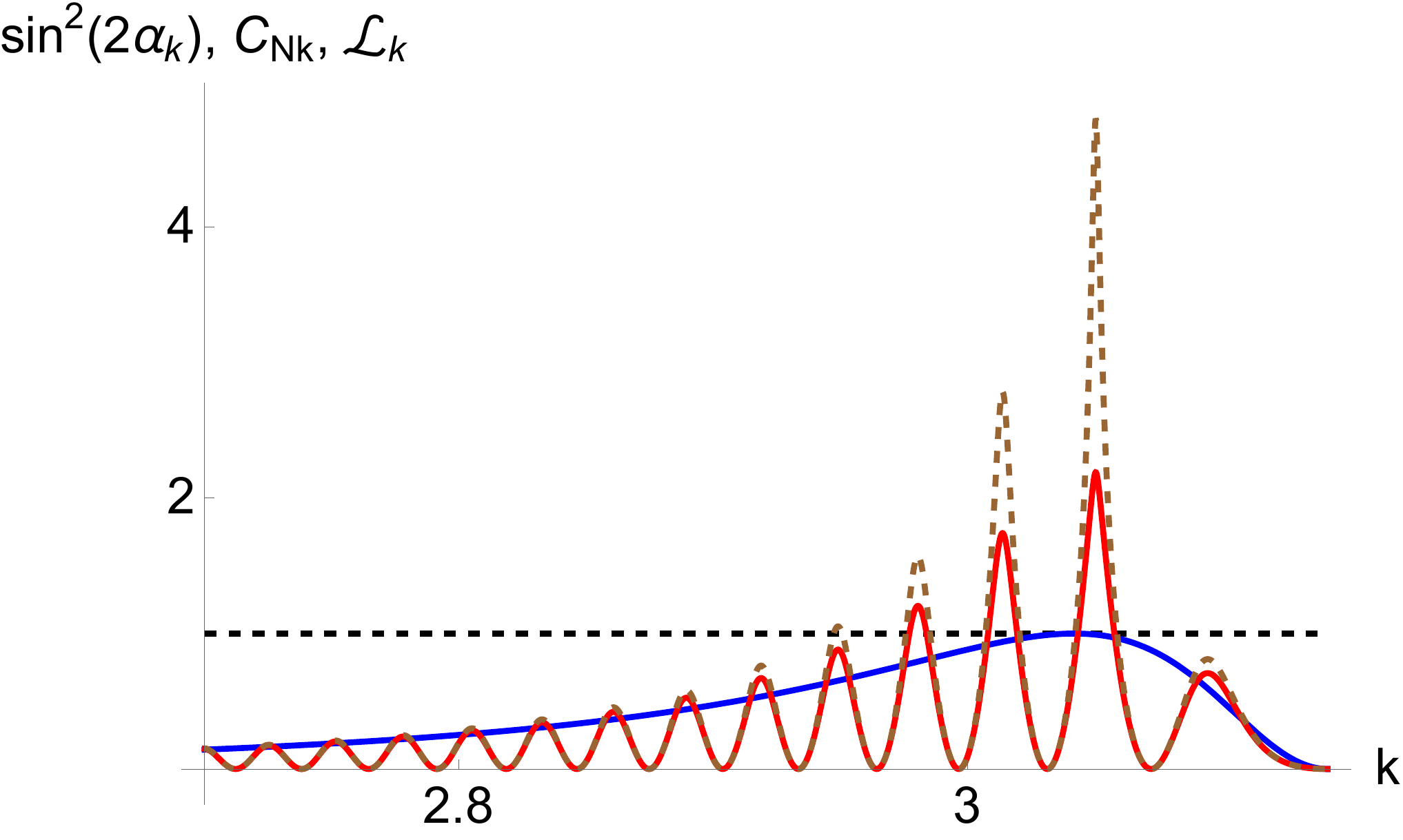}
\caption{$\sin^2(2\alpha_k)$ (solid blue), $\cnk$ at $t=200$ (dashed red) 
and $\LEk$ at $t=200$ (dashed brown)
as a function of $k$ for $h=0.95, \gamma=0.5, \delta = 0.1$. The horizontal line
marks the position of unity.}
\label{compL2}
\end{figure}

Away from the critical lines, $\sin^2(2\alpha_k) \ll 1$, and hence so is
$\sin^2(2\alpha_k)\sin^2(t\epsilon_k)$ for all times. Thus, the perturbatively
valid relation $\LE = e^{-\cn}$ continues to hold at finite times in these regions.
As one approaches $h \to 1$, this picture is more challenged, with 
the additional features being that the maximum of $\sin^2(2\alpha_k)$ shifts towards $k=\pi$, and
its maximum value approaches unity. 
This is shown in Fig. (\ref{compL2}), where the solid blue
line represents $\sin^2(2\alpha_k)$ and the solid red and dashed brown oscillating line gives $\cnk$ 
and $\LEk$ respectively, for $h=0.95, \gamma=0.5, \delta = 0.1$. In these cases, the
relation between $\cn$ and $\LE$ cannot be expresses in an exact form. 

\begin{figure}[h!]
\centering
\includegraphics[width=0.4\textwidth, height=0.3\textwidth]{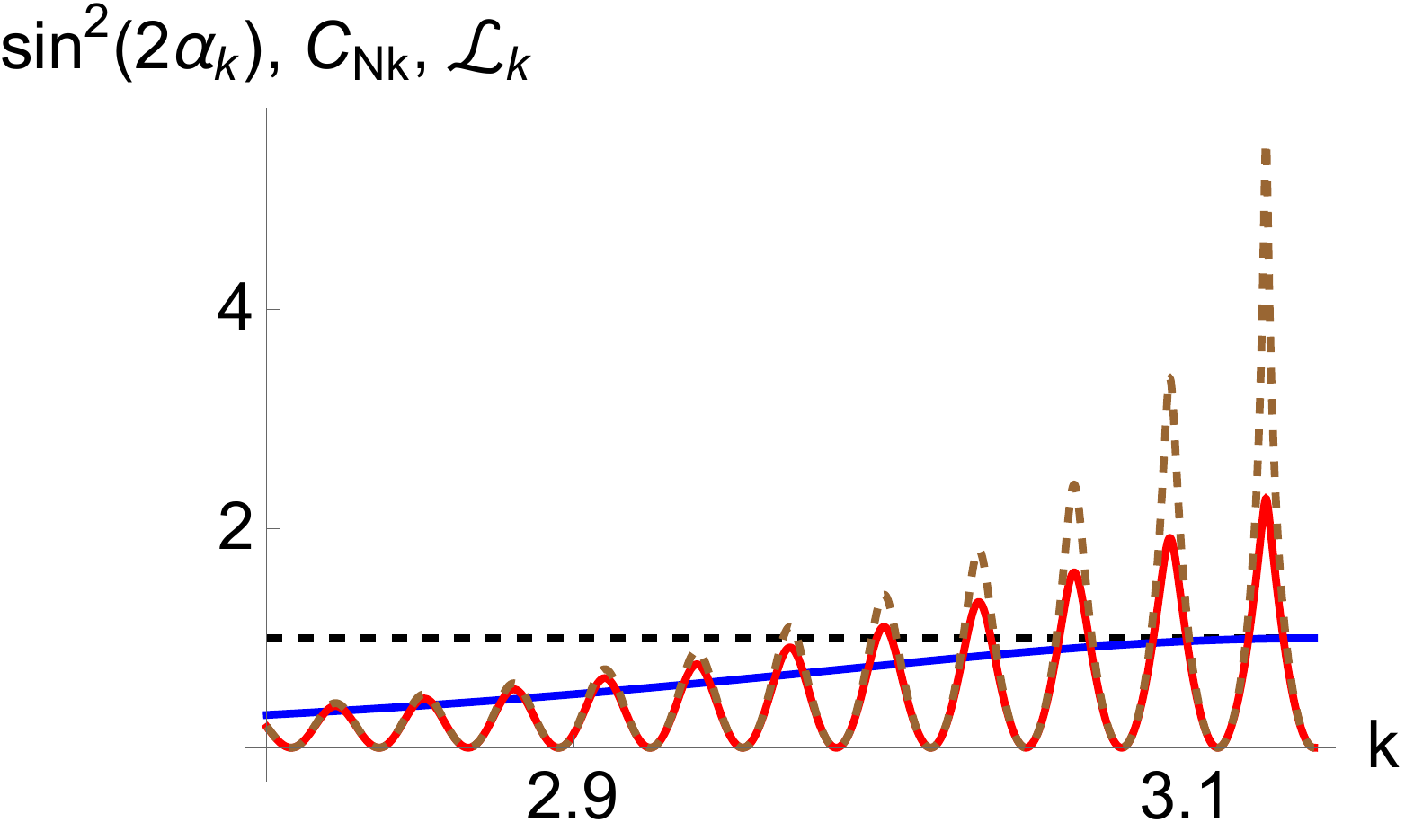}
\caption{$\sin^2(2\alpha_k)$ (solid blue), $\cnk$ at $t=200$ (solid red) 
and $\LEk$ at $t=200$ as a function of $k$ for $h=0.9, \gamma=0.5, \delta = 0.1$. 
The horizontal line marks the position of unity.}
\label{compL3}
\end{figure}

Note that when $h + \delta = 1$, the maximum of $\sin^2(2\alpha_k)\to 1$ at $k \sim \pi$.
Then, the maximally contributing mode is the one for which $k$ is close to $\pi$. 
This is the situation when after the quench the system is on the Ising critical line. 
We depict this in Fig. (\ref{compL3}), where the same color coding as Fig. (\ref{compL2}) is used.

What we further glean from the above analysis regarding the temporal dependence of $\cn$ and $\LE$
in the thermodynamic limit is the following. 
For $h$ away from the Ising critical line, for finite $t$, initially a single Fourier mode 
contributes maximally to $\cn$ and $\LE$, but as $t$ increases, more Fourier modes start contributing
to these. As a function of time, this results in the fact that while $\cn$ and $\LE$ are
initially oscillatory functions at finite time, the oscillations die out rapidly (the contributing
modes ``interfere'' destructively, as they combine from both sides of the maximum in $k$ space). 
For $h$ close to the Ising transition line, the mode
at $k\simeq\pi$ contributes maximally to $\cn$ and $\LE$.  
For $h+\delta=1$, i.e., when the quenched state is on the Ising transition line, 
$\epsilon_k \sim \pi-k \to 0$, so that 
the time period of temporal oscillations effectively become infinite. There are
thus no finite time oscillations in this case. On the contrary, when $h=1$, i.e.,
the initial state is on the Ising transition line, $\epsilon_k \to \delta$, and since
no other mode contributes significantly to $\cn$ or $\LE$, the oscillations continue for 
large times before dying down. 

\begin{figure}[h!]
\centering
\includegraphics[width=0.4\textwidth, height=0.3\textwidth]{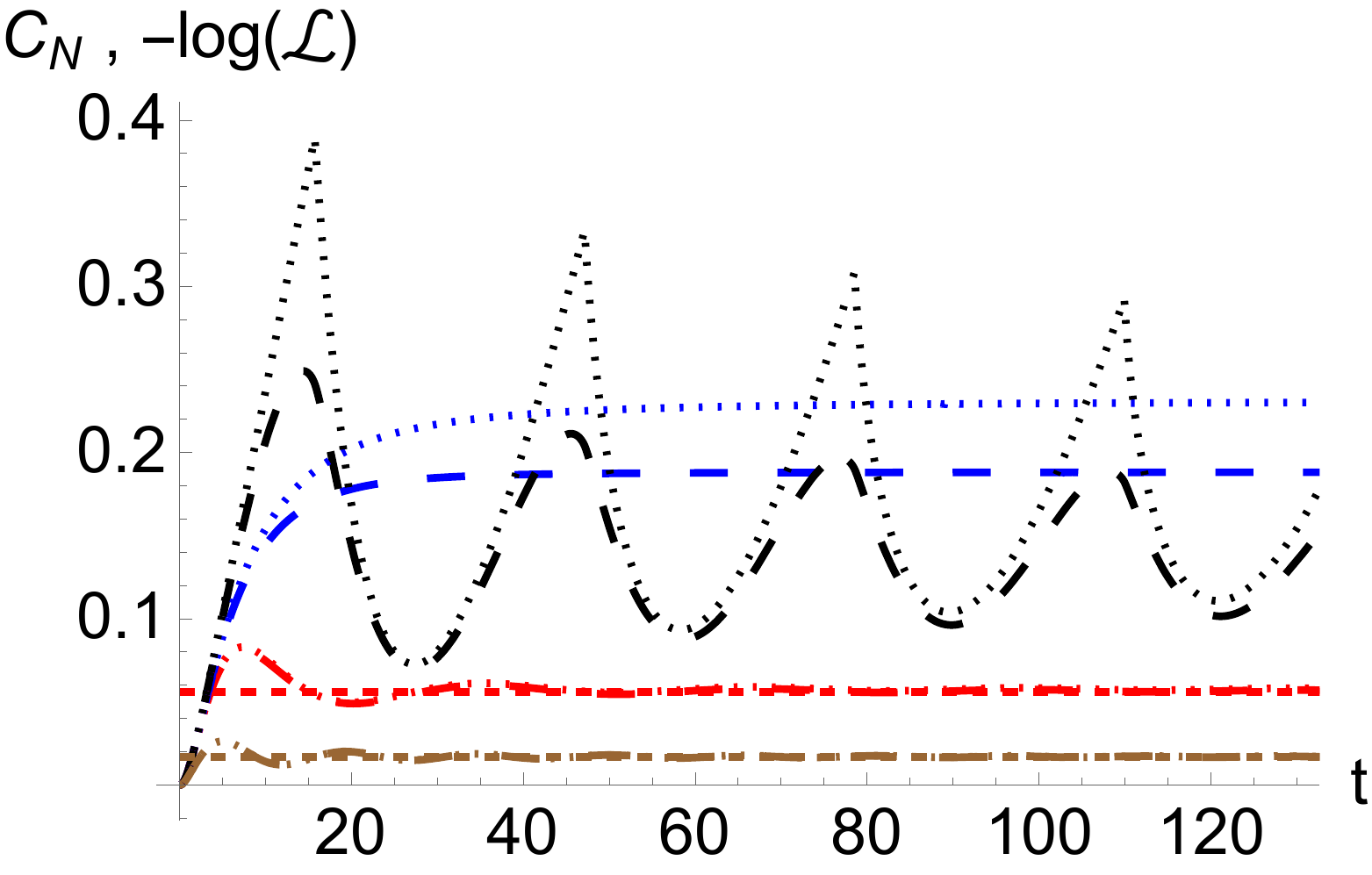}
\caption{$\cn$ (large dashed lines) and $(-\log\LE)$ (dotted lines) as a function of $t$ with $\gamma=0.5$ and $\delta=0.1$. 
The red, blue, black and brown lines are for $h=0.8$, $0.9$,
$1$ and $1.1$, respectively. The horizontal dashed red and dashed brown lines denote the
large $t$ values for $h=0.8$ and $h=1.1$, respectively.}
\label{compL4}
\end{figure}

The above arguments are only approximate, but do
capture the essential behaviour of $\cn$ and $\LE$, as shown in Fig. (\ref{compL4}) where  
we choose $\gamma=0.5$ and $\delta=0.1$. In this figure, the large-dashed red, blue, 
black and brown lines correspond to the time dependence of $\cn$ for $h=0.8$, $0.9$,
$1$ and $1.1$, respectively, while the dotted lines of the corresponding colours show these
for $\LE$. As can be seen, the curves in the figure conform to our
discussion above. Namely, for $h=0.8$ and $1.1$, there is no difference between $\cn$ and $(-\log\LE)$, and
these are indistinguishable. The difference becomes apparent when $h$ approaches unity on the ferromagnetic side. 
We have also marked by the horizontal dashed red and dashed brown
lines the large time behaviour of $\cn$, which will be discussed in the next section.
Before ending this section, we point out that there is another special case where 
the analysis is simplified, and one can obtain analytical results, namely for $h+\delta=0$. 
Here, taking $\gamma=1$ for example, it is seen that $\epsilon_k = 1$ and therefore
$\cnk$ does not have any oscillatory behaviour as a function of $k$. Hence, the temporal oscillations
do not die out in this particular case. 

\section{Transverse Quench : Complexity at large times}

At large times, the analysis of the complexity is not difficult. Our
observation here is that for such times, $\sin^2(\epsilon_k t)$ of Eq. (\ref{Phik}) becomes
a rapidly oscillating function of momentum $k$, as with increasing $t$, a very large
number of maxima (and minima) of the sin squared function can be accommodated between
$0<k<\pi$. Hence, when a large number of Fourier modes contribute to the NC, then to
a good approximation, since $\sin^2(2\Omega_k)$ is a slowly varying function, 
we can set $\sin^2(\epsilon_k t)=1/2$, i.e., its averaged value over $k$. 
This becomes challenged in two special cases, where $h=1$ and $h+\delta=1$, essentially
because a single mode contributes maximally to $\cn$ even at large $t$, and the oscillatory
behaviour of $\sin^2(\epsilon_k t)$ is less relevant there. This can already be seen in 
Fig. (\ref{compL4}), for the cases $h=0.8$ and $h=1.1$ the dashed red and dashed brown 
horizontal lines where the value of $\cn$ for the horizontal lines were computed numerically
after setting $\sin^2(\epsilon_k t)=1/2$ and one can see that these are indeed the large
time values of the NC for the corresponding values of the parameters. 

If we do not quench from or on the Ising transition lines, then to a very good approximation
we have, $\cn(t\to \infty) = \sum_k\Phi_k^2$, with $\Phi_k=\arccos(\sqrt{1-\sin^2(2\Omega_k)/2})$.
To lowest order in perturbation therefore, $\cn(t\to \infty) \sim \delta^2\sum_k(\partial\theta_{k}/\partial h)^2$, 
which is precisely $\delta^2$ times the information metric $g_{hh}$ of the time-independent case. 
We get in this case, 
\begin{eqnarray}
\cn\big|_{|h|<1} (t\to\infty)&=&\frac{\delta^2}{8|\gamma|\left(1-h^2\right)} 
+ \frac{h\delta^3}{8|\gamma|\left(1-h^2\right)^2} \nonumber\\
&-&\frac{\delta ^4 \left(13 \gamma ^2+\left(39 \gamma ^2+7\right) h^2-7\right)}{384 |\gamma|^3 \left(h^2-1\right)^3}
\nonumber\\
\cn\big|_{|h|>1} (t\to\infty)&=& \frac{\gamma ^2 \delta ^2 |h|}{8 \left(h^2-1\right) {\mathcal A}^{3}}\nonumber\\
&\mp &\frac{\gamma ^2 \delta ^3 \left(\gamma ^2+4 h^4+\left(\gamma ^2-3\right) h^2-1\right)}{16
\left(h^2-1\right)^2 {\mathcal A}^{5}}~.\nonumber\\
\label{tinf1}
\end{eqnarray}
The first few terms of the two equations above are indeed the ones we obtained in the static case,
in Eq. (\ref{stat1}), confirming our argument. 
\begin{figure}[h!]
\centering
\includegraphics[width=0.4\textwidth, height=0.3\textwidth]{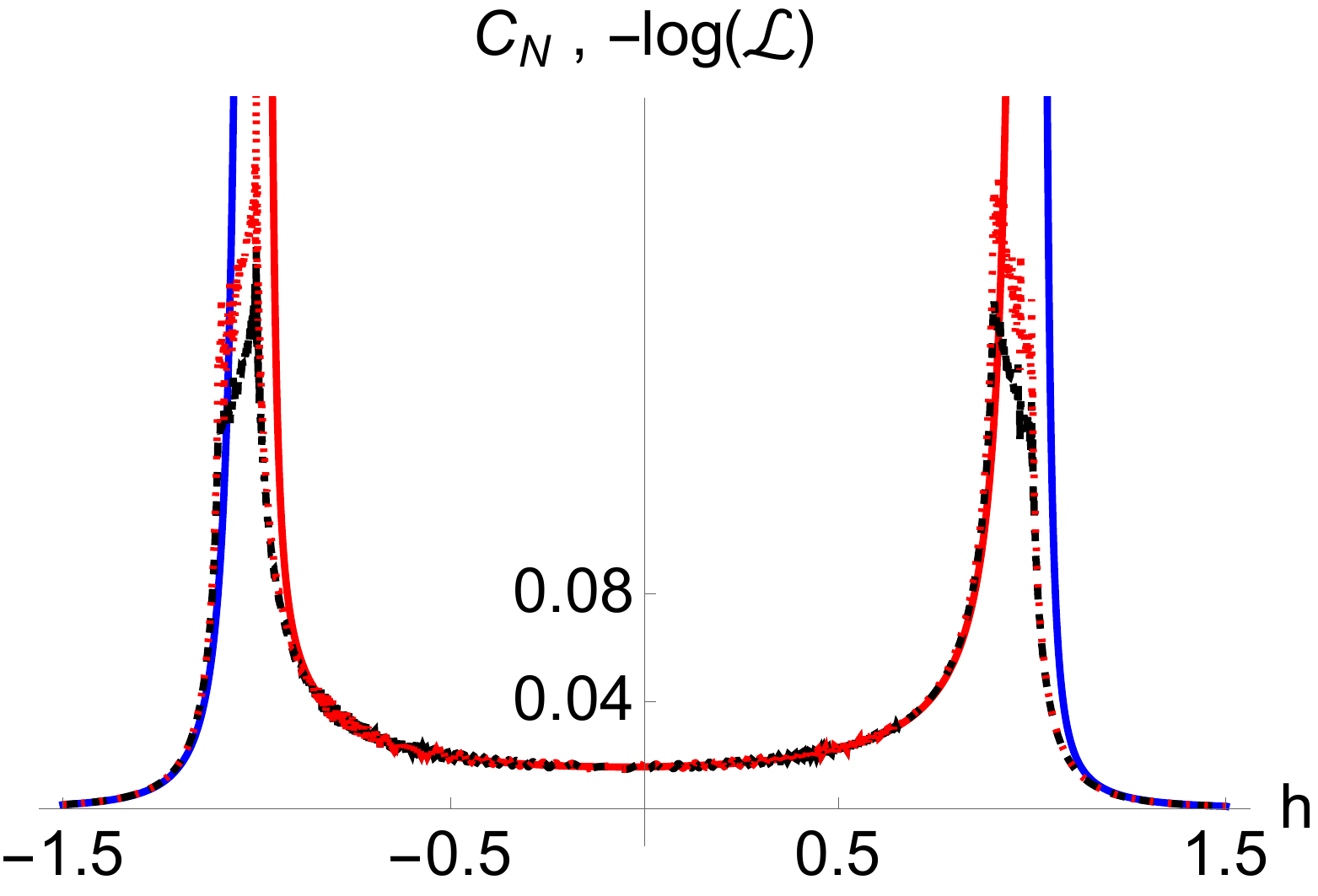}
\caption{$\cn$ and $(-\log\LE)$ as a function of $h$ with $\gamma=0.5$ and $\delta=0.1$, at large $t$. 
The dashed black curve is the numerically computed value of $\cn$ at $t=1000$, and
is indistinguishable from that of $(-\log\LE)$ shown in dotted red away from the phase boundaries. The solid 
red and blue curves are the ones computed from Eq. (\ref{tinf1}). 
}
\label{larget1}
\end{figure}
\begin{figure}[h!]
\centering
\includegraphics[width=0.4\textwidth, height=0.3\textwidth]{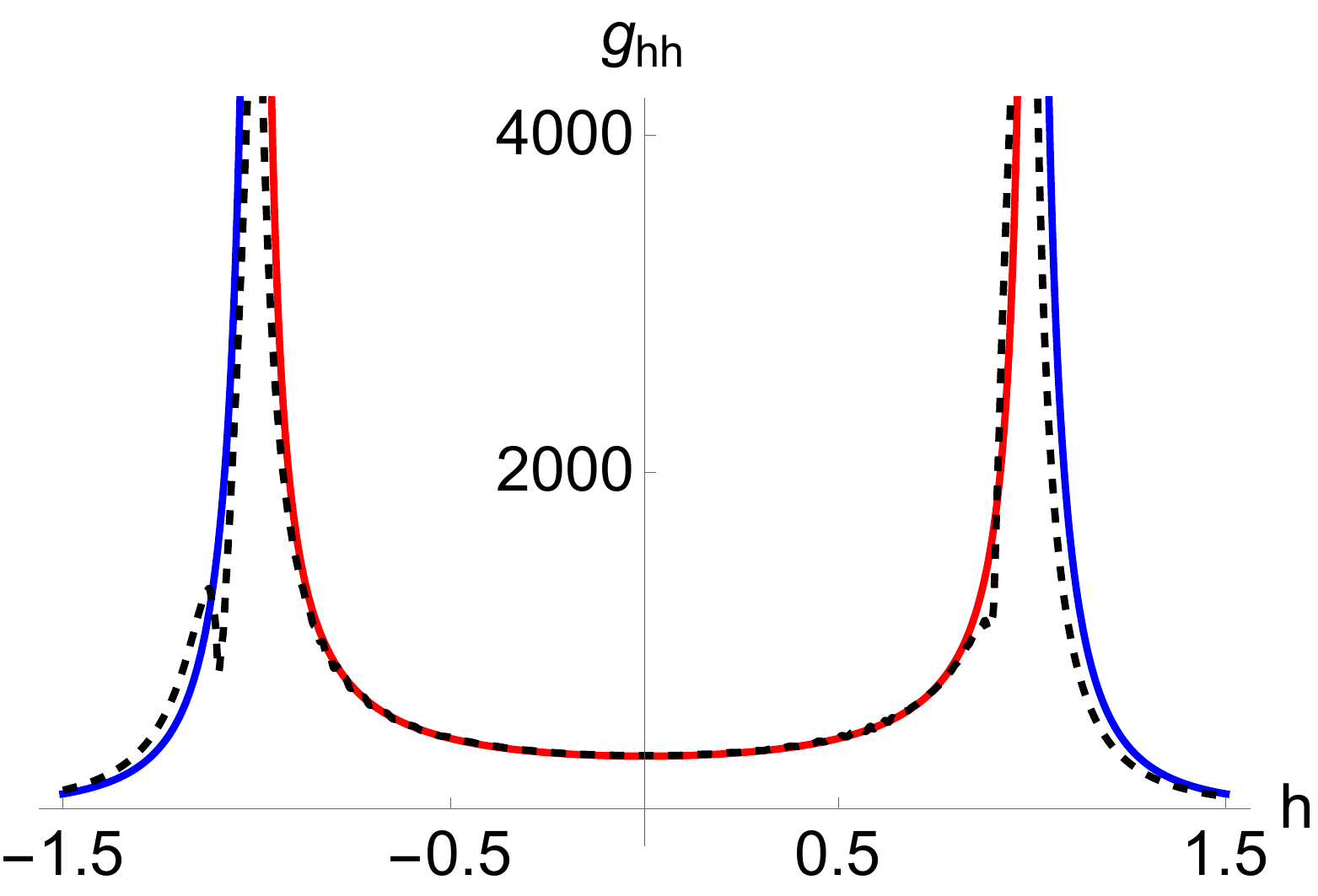}
\caption{$g_{hh}$ as a function of $h$ with $\gamma=0.5$ and $\delta=0.1$, at $t=200$. 
The solid lines are computed with Eq. (\ref{IML}) while the dashed black line shows
the result of numerical integration. 
}
\label{larget2}
\end{figure}
So the picture that emerges after a transverse quench is the following. Once the two level system is coupled to 
the transverse XY model environment, the ground state of the XY model splits and evolves in two branches. As far
as the NC is concerned, the initial evolution of the two branches give rise to completely different
structures of the complexity. However, at large times, these become identical to the static case. 
The physical reason here is not difficult to guess. At large times, the NC is essentially independent
of time, given any reference and target states, as temporal oscillations die out. Then, it is but natural that
this is similar to the static case, since there are no time scales in the problem. As an aside, we note that
the only special case where this 
will not hold is when $h+\delta=0$, as we have argued at the end of section (\ref{transversefinite}). In that
case however, the perturbative expansion in $\delta$ breaks down. 

The large time behaviour of $\cn$ and $(-\log\LE)$ is
shown in Fig. (\ref{larget1}), where the dashed black line corresponds to $\cn$ and the dotted red to 
$(-\log\LE)$, at $t=1000$, confirming our arguments above. 
The solid lines are the ones computed from Eq. (\ref{tinf1}). Here, we have chosen $\gamma=0.5$ and $\delta=0.1$.
Comparing with Fig. (\ref{compsmallt1}), we see
that starting from small times, the flat region between $h=\pm 1$ essential curves downwards so that at
large times, the shape depicted in Fig. (\ref{larget1}) is reached. We also note that close to criticality,
$(-\log\LE)$ is a few times $\cn$. The reason should be clear from Fig. (\ref{compL3}), from which
we glean that $\LE_k$ is proportional to $\mathcal{C}_{nk}$ in these regions. 

\subsection{Transverse Quench : QIM at large times}

At large times, it is straightforward to compute the IM, by analysing the expression appearing from
Eq. (\ref{wavefn}). After dropping additive terms that are highly oscillatory in
the momentum space, and average to zero, we expand the rest in powers of $\delta$. We find that 
up to ${\mathcal O}(\delta^2)$,
\begin{eqnarray}
g_{tt}\big|_{|h|<1} &=& \frac{|\gamma|\delta^2}{2\left(1+|\gamma|\right)}~,\nonumber\\
g_{hh}\big|_{|h|<1}&=& \frac{h \delta}{8 |\gamma| 
   \left(1-h^2\right)^2}\nonumber\\
   &+&\frac{\delta ^2 \left(\gamma ^2 \left(16 \left(1-h^2\right)^2
   t^2+69 h^2+23\right)+3(1- h^2)\right)}{256 |\gamma|^{3}
   \left(1-h^2\right)^3}~,\nonumber\\
g_{ht}\big|_{|h|<1}&=&0~,
\label{IML}
\end{eqnarray}
with the corresponding expressions for the region $|h|>1$ being too lengthy to reproduce here. 
\begin{figure}[h!]
\centering
\includegraphics[width=0.4\textwidth, height=0.3\textwidth]{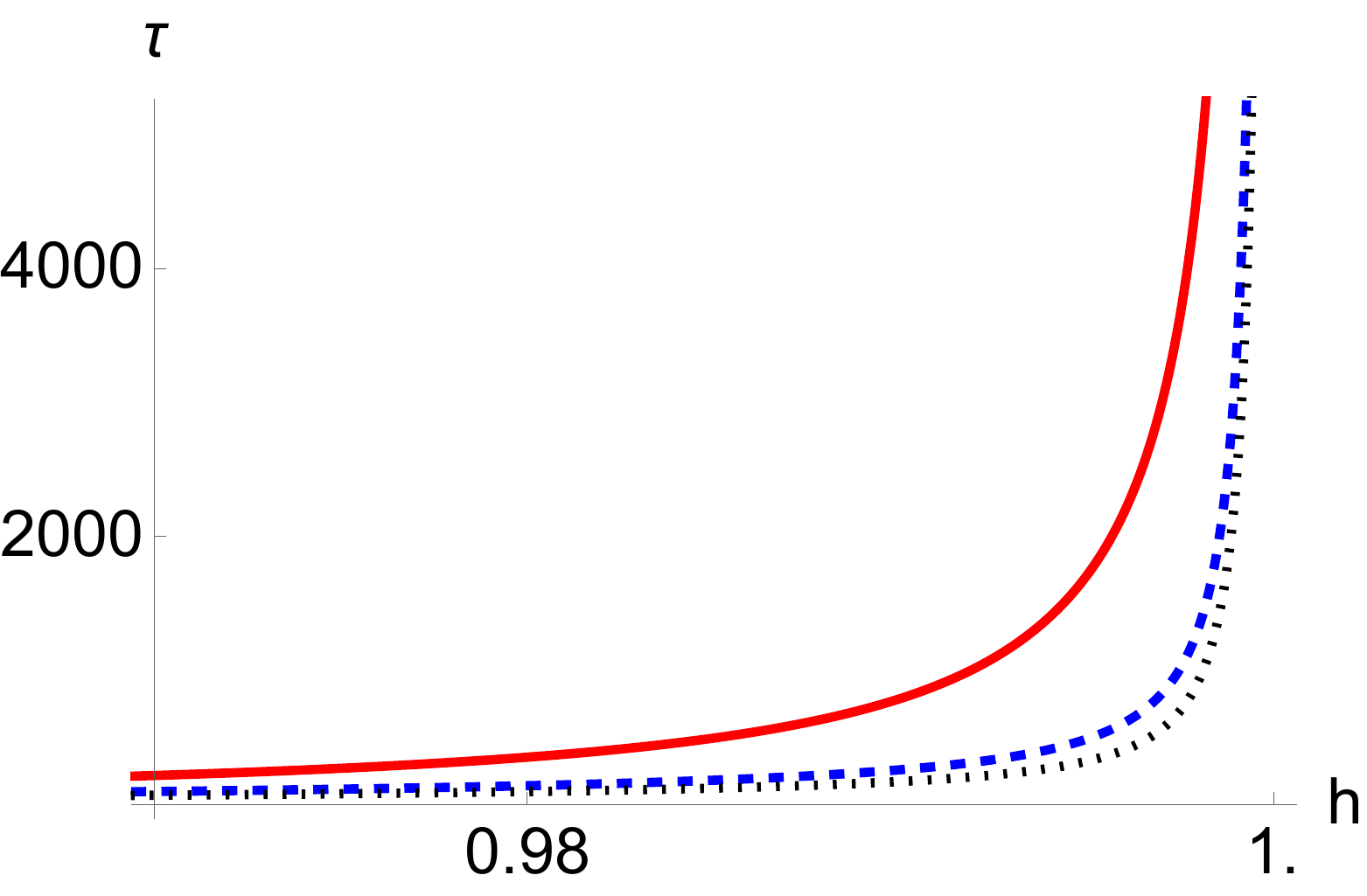}
\caption{$\tau$ as a function of $h$ with $\gamma=0.5$ and $\delta=0.1$, at $t=200$,
computed from Eq. (\ref{IML}). Here, the solid red, dashed blue and dotted black lines correspond
to the initial values of $h=0.8$, $0.85$ and $0.9$, respectively.}
\label{larget3}
\end{figure}

Unlike the NC, the QIM at large times is different from the static case derived in \cite{Zanardi}, \cite{Polkov},
although curiously at the lowest order, $g_{hh}dh^2$ has the same form as the first correction to the
ground state complexity of Eq. (\ref{stat1}), upon identifying $dh \sim \delta$. Importantly, here the $t-t$
component of the metric on the branch coupled to the excited state of the central spin half system remains
the same as in the limit of small times, and hence at large times the QIM does not reduce to the static
situation. 

One can now compute the geodesics on this parameter manifold, as discussed in 
subsection \ref{fscsmallt}. We depict this pictorially in Fig. (\ref{larget3}), where we have
chosen $\gamma=0.5$, $\delta=0.1$, at the starting time is $t=200$. 
In this figure, the solid red, dashed blue and dotted black lines correspond
to the initial values of $h=0.8$, $0.85$ and $0.9$, respectively.
We see that the FSC 
exhibits expected behaviour here, i.e., the derivative of $\tau$ diverges near the critical line
$h=1$. For $t\to \infty$, this divergence is $\sim (1-h)^{-3/2}$. 

\section{Scaling relations for the NC}

From our discussion in section (\ref{transversefinite}), it should be clear that almost all
results that have been derived in the literature regarding the finite size scaling relations
of $\LE$ (see, e.g., \cite{LE3} and references therein) will continue to remain valid for
the exponential of $(-\cn)$. Hence a separate analysis of the finite size behaviour of 
$\cn$ is not necessary. For example, for both for large and small times, 
\begin{equation}
\frac{\partial\cn(t)}{\partial\lambda}\sim N~.
\end{equation}
with $\lambda = h,\gamma$. For completeness, we will record the behaviour of the derivative of
$\cn$ with respect to the system parameters. For small times, we get
\begin{equation}
\frac{1}{N}\frac{\partial\cn(t)}{\partial(h+\delta)}\sim |h+\delta\pm1|~,~~
\frac{1}{N}\frac{\partial\cn(t)}{\partial(\gamma+\delta)}\sim |\gamma+\delta|~,
\end{equation}
near the Ising transition line and the anisotropic transition line, respectively. 
For large times, these relations change to 
\begin{equation}
\frac{1}{N}\frac{\partial\cn(t\rightarrow\infty)}{\partial(h+\delta)}\sim \log|h+\delta\pm 1|~,~~
\frac{1}{N}\frac{\partial\cn(t\rightarrow\infty)}{\partial(\gamma+\delta)}\sim \log|\gamma+\delta|~.
\end{equation}

\section{Conclusions}

The Nielsen complexity, the Fubini-Study complexity and the Loschmidt echo are three fundamental
quantities of interest in studies on quantum criticality. 
In this paper, we have performed a detailed analysis of these quantities for the transverse XY model
in the presence of a sudden quantum quench, in the thermodynamic limit. The complexities were considered 
both in a static scenario as well as one involving the
quench, and we compared the latter to the LE. While the NC in a static scenario was 
considered by us earlier in \cite{tapo1}, here we have
computed analytical expressions for the NC and the FSC both in the static as well as the
quench scenarios in a perturbative setup, in the small and large time limits. 
This also gives analytical expressions for the LE, from what we have discussed. 

We have shown that at small times, these quantities are related by $\LE = e^{-\cn} \sim$ $e^{-d\tau^2}$. 
The simple relation between the three physical quantities at small times is indeed remarkable.  
However, although the first relation here continues to hold at large times, the second one does not.
To wit, the evolution of the transverse XY chain proceeds in two distinct branches, in which the
ground state of the model is coupled to the ground and the excited states of the central spin. 
As far as the NC is concerned, after a long time, temporal oscillations die down and the 
NC reduces to the static case, as we have argued. The QIM, on the other hand, evolves differently,
and its large time behaviour bears little resemblance to the static scenario, in the presence
of a time component of the metric. 

Note that in this paper, we have considered only the transverse field XY model. Now, as 
discussed in \cite{tapo1}, general features of the analysis here should be applicable to all 
quadratic Hamiltonians, as these only depend on the Bogoliubov angle. We end by commenting that 
in quench scenarios, the triangle inequality associated to the NC seems to be violated for 
both small and finite times, as we have checked both analytically and numerically. 
The exact implication of this is unclear to us as of now, and this warrants further study. \\

\noindent
{\bf Acknowledgments}\\

\noindent
N. J. would like to acknowledge the University Grant Commission (UGC), India, 
for providing financial support. The work of T. S. is supported in part by Science and Engineering
Research Board (India) via Project No. EMR/2016/008037.

\appendix

\section{}
\label{AppendixA}

In this appendix, we will list the computation of the NC and the IM for an anisotropic quench, at
small times. The results for the region $|h|>1$ become lengthy and we will only present those 
in the region $|h|<1$. Here, we find the lowest order terms 
\begin{eqnarray}
\cn\big|_{|h|<1} &=& \cn^{reg}+ \frac{ \delta ^2 t^2 \left(|\gamma| +2 |\gamma|  h^2+1\right)}{4 (|\gamma| +1)^3}
\mp\frac{|\gamma|  \delta ^3 t^4 \left(3+ |\gamma| {\mathcal B}\right)}{24 (|\gamma| +1)^5}~,~\nonumber\\
{\mathcal B} &=&|\gamma|  (|\gamma| +5)-8 h^4+4 (|\gamma| +1) (|\gamma| +4) h^2+7~,~\nonumber\\
\cn^{reg}&=&-\frac{1}{48} \delta ^2 \left(4h^2+1 \right) t^4~,
\label{aniso1}
\end{eqnarray}
where the $-(+)$ sign is for $\gamma > 0 (<0)$, respectively. 
The IM is given to lowest order by
\begin{eqnarray}
g_{tt}&=&\frac{\delta ^2 \left(|\gamma| +2 |\gamma|  h^2+1\right)}{4 (|\gamma| +1)^3}~,~
g_{\gamma\gamma}=\mp\frac{t^4|\gamma|\delta\left(3+|\gamma|{\mathcal B}\right)}{24 (|\gamma| +1)^5}~,~\nonumber\\
g_{t\gamma}&=&\frac{t}{\delta}g_{tt}+\frac{\delta}{t}g_{\gamma\gamma}-\frac{1}{24} \delta  \left(4 h^2+1\right) t^3~.
\end{eqnarray}

\end{document}